\documentclass[12pt]{article}

\usepackage{latexsym}
\usepackage{amssymb,amsfonts,amsmath}
\usepackage{graphicx}
\usepackage{indentfirst}
\usepackage{bbm}
\usepackage{amssymb}
\usepackage{verbatim}
\usepackage{amsmath, amsthm,amssymb}
\usepackage{mathrsfs}
\usepackage{hyperref}
\usepackage{amsfonts}
\usepackage{dsfont}
\usepackage{cite}
\usepackage{xcolor}
\usepackage[open,openlevel=2,atend,numbered]{bookmark}
\usepackage[multiple]{footmisc}
\parskip=\medskipamount

\counterwithin*{equation}{section}

\newcommand{\be}{\begin{equation}}
\newcommand{\ee}{\end{equation}}
\newcommand{\bea}{\begin{eqnarray}}
\newcommand{\eea}{\end{eqnarray}}
\newcommand{\non}{\nonumber}
\newcommand {\cD}{{\cal D}}

\newcommand {\cH}{{\cal H}}

\newcommand {\cL}{{\cal L}}

\newcommand {\cN}{{\cal N}}

\newcommand {\cP}{{\cal P}}

\newcommand {\cW}{{\cal W}}
\newcommand {\cX}{{\cal X}}
\newcommand {\cY}{{\cal Y}}
\newcommand {\cZ}{{\cal Z}}

\renewcommand{\a}{\alpha}
\renewcommand{\b}{\beta}
\renewcommand{\c}{\chi}
\renewcommand{\d}{\delta}
\newcommand{\e}{\epsilon}

\newcommand{\f}{\phi}

\newcommand{\g}{\gamma}

\newcommand{\m}{\mu}
\newcommand{\n}{\nu}

\renewcommand{\r}{\rho}
\newcommand{\s}{\sigma}

\newcommand{\w}{\omega}

\newcommand{\D}{\Delta}

\newcommand{\G}{\Gamma}
\renewcommand{\L}{\Lambda}

\newcommand{\U}{\Upsilon}

\newcommand{\X}{\Xi}

\newcommand{\Tr}{\text{Tr}}
\newcommand{\tr}{\text{Tr}}
\newcommand{\rT}{\text{T}}
\newcommand{\rd}{\text{d}}
\newcommand{\ri}{\text{i}}
\newcommand{\re}{\text{e}}

\newcommand{\vf}{\varphi}

\newcommand{\fb}{\bar{\phi}}
\newcommand{\ib}{\bar{i}}
\newcommand{\jb}{\bar{j}}
\newcommand{\kb}{\bar{k}}
\newcommand{\lb}{\bar{l}}
\newcommand{\Ib}{\bar{I}}
\newcommand{\Jb}{\bar{J}}
\newcommand{\Kb}{\bar{K}}
\newcommand{\Lb}{\bar{L}}

\newcommand{\ad}{\dot{\alpha}}
\newcommand{\bd}{\dot{\beta}}

\newcommand{\tF}{\tilde{F}}

\newcommand{\hm}{\hat{m}}
\newcommand{\hn}{\hat{n}}
\newcommand{\hp}{\hat{p}}
\newcommand{\hq}{\hat{q}}
\newcommand{\hr}{\hat{r}}
\newcommand{\hs}{\hat{s}}

\newcommand{\dsR}{{\mathbb R}}

\newcommand{\sSp}{\mathsf{Sp}}
\newcommand{\sSU}{\mathsf{SU}}
\newcommand{\sSL}{\mathsf{SL}}
\newcommand{\sGL}{\mathsf{GL}}
\newcommand{\sSO}{\mathsf{SO}}

\newcommand{\sU}{\mathsf{U}}

\newcommand{\sMat}{\mathsf{Mat}}

\newcommand{\ph}{\phantom}
\newcommand{\id}{\mathds{1}}
\renewcommand{\non}{\nonumber}

\newcommand{\pd}{\partial}
\renewcommand{\~}{\tilde}

\newcommand{\nab}{\nabla}

\newcommand{\inte}{\int\!\!\rd}

\newcommand{\ve}{\varepsilon}                            %new
\newcommand{\pa}{\partial}                           %new
\newcommand{\hf}{\frac12}
\begin{document}

%%%%
\begin{titlepage}
\begin{flushright}
January, 2023\\
\end{flushright}
\vspace{5mm}

\begin{center}
{\Large \bf Weyl invariance, non-compact duality and conformal higher-derivative sigma models}
\end{center}

\begin{center}

{\bf

Darren T. Grasso, Sergei M. Kuzenko and Joshua R. Pinelli}
\vspace{5mm}

\footnotesize{
{\it Department of Physics M013, The University of Western Australia\\
35 Stirling Highway, Crawley W.A. 6009, Australia}}
\vspace{2mm}
~\\
Email: \texttt{darren.grasso@uwa.edu.au, sergei.kuzenko@uwa.edu.au, joshua.pinelli@research.uwa.edu.au}\\
\vspace{2mm}

\end{center}

\begin{abstract}
\baselineskip=14pt
We study a system of $n$ Abelian vector fields coupled to $\frac 12 n(n+1)$ complex scalars parametrising the Hermitian symmetric space
$\mathsf{Sp}(2n, {\mathbb R})/ \mathsf{U}(n)$. This model is Weyl invariant and possesses the maximal non-compact duality group $\mathsf{Sp}(2n, {\mathbb R})$.
Although both symmetries are anomalous in the quantum theory, they should be respected by the logarithmic divergent term (the ``induced action'') of the effective action obtained by integrating out the vector fields.
We compute this induced action and demonstrate its Weyl and $\mathsf{Sp}(2n, {\mathbb R})$ invariance. The resulting conformal higher-derivative $\sigma$-model on $\mathsf{Sp}(2n, {\mathbb R})/ \mathsf{U}(n)$ is generalised to the cases where the fields take their values in (i) an arbitrary K\"ahler space; and (ii) an arbitrary Riemannian manifold. In both cases, the $\sigma$-model Lagrangian generates a Weyl anomaly satisfying the Wess-Zumino consistency condition.
\end{abstract}

\vfill

\vfill
\end{titlepage}

\newpage
\renewcommand{\thefootnote}{\arabic{footnote}}
\setcounter{footnote}{0}

\tableofcontents{}
\vspace{1cm}
\bigskip\hrule

\allowdisplaybreaks

%%%%%%%%%%%%%%%%%%%%%%%%%%%%%%%%

\section{Introduction}

A unique feature of the Weyl multiplet of $\cN=4$ conformal supergravity \cite{BdeRdeW} is the presence of a dimensionless complex scalar field $\f$ that parametrises the Hermitian symmetric space
$\sSL(2,{\mathbb R}) /\sSO(2)$.\footnote{This coset space can equivalently be realised as $\sSU(1,1) /\sU(1)$, since the groups $\sSL(2,{\mathbb R})$ and $\sSU(1,1)$ are isomorphic. Refs. \cite{BCdeWS,BCS} make use of the latter group.}
The most general family of invariant actions for $\cN=4$ conformal supergravity was derived only a few years ago by Butter, Ciceri, de Wit and Sahoo,
\cite{BCdeWS,BCS}.
Such an action is uniquely determined by a holomorphic function $\cH(\f)$ which accompanies the terms quadratic in the Weyl tensor in the Lagrangian.

 For the special choice $\cH = {\rm const}$, in which case the $\cN=4$ conformal supergravity action proves to be invariant under rigid  $\sSL(2,{\mathbb R}) $ transformations, the corresponding action
 was constructed in 2015 by Ciceri and Sahoo \cite{CS} to second order in fermions. The bosonic sector of the latter action had been computed in 2012 by Buchbinder, Pletnev and Tseytlin \cite{BPT}  as an ``induced action'', obtained by integrating out an Abelian $\cN=4$ vector multiplet coupled to external $\cN= 4 $ conformal
 supergravity.\footnote{Some
of the relevant terms were missed in \cite{BPT}.}
The purely $\f$-dependent part of the Lagrangian is a higher-derivative
$\s$-model of the form \cite{Osborn}:
\bea
\cL (\f, \bar \f) &=&  \frac{1}{({\rm Im}\, \f)^2} \Big[ \cD^2 \f \cD^2 \bar \f
- 2 (R^{mn}- \frac 13 g^{mn} R) \nabla_m \f \nabla_n \bar \f \Big] \non \\
&&+ \frac{1}{12 ( {\rm Im}\, \f)^4}
\Big[ \a \nabla^m \f \nabla_m \f \nabla^n \bar \f  \nabla_n \bar \f
+\b \nabla^m \f \nabla_m \bar  \f \nabla^n  \f  \nabla_n \bar \f
\Big] ~,
\label{1.1}
\eea
where
\bea
\cD^2 \f := \nabla^m \nabla_m \f + \frac{\ri}{{\rm Im}\, \f} \nabla^m \f \nabla_m \f ~,
\label{1.2}
\eea
and $\a$ and $\b$ are numerical parameters. In the case of $\cN=4$ conformal supergravity, these coefficients are \cite{BPT}: $\a = \hf \b =1$.
The Lagrangian \eqref{1.1} is invariant under $\sSL(2,{\mathbb R}) $
transformations
\bea
\f \to \f' = \frac{a\f + b}{c\f +d} ~, \qquad \left(
\begin{array}{cc}
 a  & \quad b\\
c &  \quad d
\end{array}
\right) \in \sSL(2,{\mathbb R})
\label{1.3}
\eea
acting on the upper half-plane ${\rm Im}\, \f >0$ with metric
\bea
\rd {\mathfrak s}^2 = \frac{ \rd \f \,\rd \bar \f}{ ( {\rm Im}\, \f)^2 } \quad \Longrightarrow \quad
\G^\f_{\f \f} = \frac{\ri}{{\rm Im}\, \f} ~, \quad
\G^{\bar \f}_{\bar \f \bar \f} = - \frac{\ri}{{\rm Im}\, \f} ~.
\eea
The functional $\int \rd^4 x \, \sqrt{-g}\, \cL $
proves to be invariant under Weyl transformations
\bea
g_{mn} (x) \rightarrow \re^{2\s (x) }g_{mn}(x) ~,
\label{Weyl}
\eea
since the scalar field $\f$ is inert under such transformations.
The higher-derivative $\s$-model \eqref{1.1} possesses the $\cN=1$ supersymmetric extension  \cite{K2020} which relates the parameters $\a$ and $\b$. Both parameters are completely fixed if $\cN=2$ supersymmetry is required \cite{K2020, deWit:2010za, GHKSST}.

The conformal higher-derivative $\s$-model \eqref{1.1} admits a nontrivial generalisation
that is obtained by replacing the Hermitian symmetric space
$\sSL(2,{\mathbb R}) / \sSO(2) $ with an arbitrary
$n$-dimensional K\"ahler manifold  ${\mathfrak M}^{n}$, with $n$ the complex dimension.
We assume that ${\mathfrak M}^{n}$ is
parametrised by $n$ local complex coordinates $\f^I$ and their conjugates $\bar \f^{\Ib}$. Let ${\mathfrak K}(\f, \bar \f)$ be the corresponding
K\"ahler potential such that the K\"ahler metric ${\mathfrak g}_{I\Jb}(\f,\fb)$ is given by
${\mathfrak g}_{I\bar J} = \pa_I \pa_{\bar J} {\mathfrak K}$.
Associated with ${\mathfrak M}^n$ is a  higher-derivative sigma model
of the form
\begin{align}
	S=\inte^4x\,\sqrt{-g} \,&\bigg\{
		{\mathfrak g}_{I\Jb}(\f,\fb)\Big[\cD^2\f^{I}\cD^2\fb^{\Jb}
	-2\Big(R^{mn}-\frac{1}{3}Rg^{mn}\Big)\nab_m\f^{I}\nab_n\fb^{\Jb}\Big] \non \\
	&
+ {\mathfrak F}_{IJ \Kb \Lb} (\f,\fb)	
\nab^m\f^{I} \nab_m\f^{J}  \nab^n\fb^{\Kb} \nab_n\fb^{\Lb} \non \\
& +\Big[ {\mathfrak G}_{IJ \Kb \Lb} (\f,\fb)	
\nab^m\f^{I} \nab^n\f^{J}  \nab_m\fb^{\Kb} \nab_n\fb^{\Lb} +{\rm c.c.} \Big]
\label{1.4}
\bigg\} ~,
\end{align}
where $R_{mn}$ is the spacetime Ricci tensor,
\begin{align}
	\cD^2\f^I:=\nabla^m \nabla_m \f^I+\G^I_{\ph{I}JK}(\f, \bar \f)
	\nab^m\f^J\nab_m\f^K~,
	\end{align}
with $\G^I_{\ph{I}JK}$ being the Christoffel symbols for the K\"ahler metric ${\mathfrak g}_{I\Jb}$.
Finally,
${\mathfrak F}_{IJ \Kb \Lb} $ and ${\mathfrak G}_{IJ \Kb \Lb} $ are tensor fields on the
target space, which are constructed from the K\"ahler metric ${\mathfrak g}_{I\Jb}$,
Riemann tensor ${\mathfrak R}_{I \Jb K \Lb} $ and, in general, its covariant derivatives.
We recall that the Christoffel symbols $\G^I_{\ph{I}JK} $ and the curvature tensor ${\mathfrak R}_{I \bar J K  \bar L} $ are given by the expressions\footnote{The reader is referred, e.g., to \cite{WB,Buchbinder:1998qv,FVP} for a pedagogical review of K\"ahler geometry in the framework of nonlinear sigma models.}
\bea
\G^I_{\ph{I}JK} = {\mathfrak g}^{I \bar L} \pa_J \pa_K \pa_{\bar L} {\mathfrak K}~,\qquad
 {\mathfrak R}_{I \bar J K  \bar L} = \pa_I \pa_K \pa_{\bar J} \pa_{\bar L}  {\mathfrak K}
-{\mathfrak g}^{M \bar N} \pa_I \pa_K \pa_{\bar N} {\mathfrak K}  \pa_{\bar J} \pa_{\bar L}  \pa_M {\mathfrak K}~.
\label{18}
\eea
A typical expression for $ {\mathfrak F}_{IJ \Kb \Lb}$ is
\bea
 {\mathfrak F}_{IJ \Kb \Lb} =  \a_1 {\mathfrak R}_{(I \Kb J)\Lb}
 +\a_2 {\mathfrak g}_{(I \Kb} {\mathfrak g}_{J) \Lb} +\dots
 \label{F-structure}
 \eea
The possible structure of ${\mathfrak G}_{IJ \Kb \Lb}$ is analogous.
It should be pointed out that actions of the form \eqref{1.4} naturally emerge at the component level in $\cN=2$ superconformal higher-derivative $\s$-models
\cite{deWit:2010za} (see also \cite{GHKSST}),  and in $\cN=1$ ones \cite{K2020}.

By construction, the action \eqref{1.4} is invariant under arbitrary holomorphic isometries of  ${\mathfrak M}^{n}$. A nontrivial observation is that \eqref{1.4} is also invariant under arbitrary Weyl transformations of spacetime provided the scalars $\f^I$ are inert under these transformations.
Choosing  ${\mathfrak M}^{n}={\mathbb C}^n$ and ${\mathfrak g}_{I\Jb}(\f,\fb) = \d_{I\Jb}$
in \eqref{1.4} and integrating by parts, one obtains
 the Fradkin-Tseytlin (FT)
operator \cite{FT1982}
\bea
\Delta_0 = (\nabla^m \nabla_m)^2 + 2 \nabla^m \big(
	 {R}_{mn} \,\nabla^n
	- \frac{1}{3} {R} \,\nabla_m
	\big)~,
\eea
which is conformal when acting on dimensionless scalar fields.\footnote{This operator was  re-discovered
by Paneitz in 1983 \cite{Paneitz} and Riegert in 1984 \cite{Riegert}.}
Given a Weyl inert scalar field $\vf$, the Weyl transformation \eqref{Weyl} acts on
$\D_0 \vf$ as
\bea
\D_0 \vf~ \to ~\D_0 \vf = \re^{-4\s} \D_0 \vf~.
\eea

 An action of the form $\int \rd^4 x \, \sqrt{-g}\, \cL (\f, \bar \f)$, with $\cL$ given by \eqref{1.1}, naturally emerges as an induced action in Maxwell's electrodynamics coupled to a dilaton
 $\vf$ and an axion  $\mathfrak a$ with Lagrangian
\bea
L(F; \f , \bar \f) &=& - \frac 14 \re^{-\vf} F^{mn}F_{mn} -\frac 14 {\mathfrak a} \tilde F^{mn} F_{mn} \non \\
&=&  \frac{\ri}{2} \f F^{\a\b} F_{\a\b} + {\rm c.c.}
~,
\qquad \quad \f = {\mathfrak a} +\ri e^{-\vf}~.~~
\label{1.8}
\eea
Here $\tilde F^{mn}= \hf \ve^{mnrs} F_{rs}$ is the Hodge dual of the electromagnetic field strength
$F_{mn}=2\nabla_{[m} A_{n]} = 2\pa_{[m} A_{n]} $, with $\ve^{mnrs}$ the Levi-Civita tensor.
The second form of the Lagrangian \eqref{1.8} is written using two-component spinor notation, where the field strength $F_{mn} = - F_{nm}$ is replaced with a symmetric rank-2 spinor $F_{\a\b} = F_{\b\a}$ and its conjugate $\bar F_{\ad \bd}$.
More precisely, if one considers the effective action, $\G [\f, \bar \f]$,
obtained by integrating out the quantum gauge field
in the model \eqref{1.8}, then the logarithmically divergent part of $\G [\f, \bar \f]$
is given by $\int \rd^4 x \, \sqrt{-g}\, \cL (\f, \bar \f)$, as demonstrated by Osborn \cite{Osborn}.
An important question arises: why is the induced action Weyl and  $\sSL(2,{\mathbb R}) $ invariant?

We recall that  the group of electromagnetic duality rotations of free Maxwell's equations is  $\sU(1)$.
More than forty years ago, it was shown by Gaillard and Zumino \cite{GZ1,Zumino} that
the non-compact group $\sSp(2n, {\mathbb R})$ is the maximal duality group of
$n$ Abelian vector field strengths $F_{mn}=  (F_{mn, i})$, with $i =1, \dots, n$, in the presence of a collection of complex scalars $\f^{ij}=\f^{ji}$ parametrising the homogeneous
space $\sSp(2n, {\mathbb R})/\sU(n)$, with $i, j = 1, \dots , n$.
In the absence of such scalars, the largest duality group proves to be
$\sU(n)$, the maximal compact subgroup of $\sSp(2n, {\mathbb R})$.
These results admit a natural extension to the case when the pure vector field
part  $L(F)$ of the Lagrangian $L(F;\f, \bar \f)$ is a nonlinear $\sU(1)$ duality invariant  theory
 \cite{GR1,GR2,GZ2,GZ3,AT} (see \cite{KT2,AFZ,Tanii} for reviews), for instance
  Born-Infeld theory. However, in the case that $L(F)$ is quadratic, the $F$-dependent part of $L(F; \f, \bar \f)$ is also invariant under the Weyl transformations in curved space. Then, computing the path integral over the gauge fields  leads
 to an effective action, $\G [\f, \bar \f]$, such that its logarithmically divergent part
 is invariant under Weyl and
rigid $\sSp(2n, {\mathbb R})$ transformations, see, e.g., \cite{FT,RT} for formal arguments. Both symmetries are anomalous at the quantum level, but the logarithmically divergent part of the one-loop effective action is invariant under these transformations.

In this paper we demonstrate that an action of the type \eqref{1.4} emerges as an induced action in a model for $n$ Abelian gauge fields $A_{m}=\left(A_{m,i}\right)$, $i=1,\ldots,n$, coupled to a complex field $\f=(\f^{ij}) $
and its conjugate $\fb=(\fb^{\,\ib\jb})$
parametrising the homogeneous space $\sSp(2n,\dsR)/\sU(n)$,
\begin{equation}
  \f=\f ^{\rT}\in \sMat (n, {\mathbb C})
	~,\qquad \ri(\fb-\f)>0~.\label{eqn:DilatonAxionConstraints}
\end{equation}
The corresponding Lagrangian is
\bea
L^{(n)}(F;\f,\fb)&=&  -\frac{1}{4}
\Big\{\left(F^{mn}\right)^\rT \X F_{mn}
	+\left(F^{mn}\right)^\rT \U \tF_{mn}\Big\} \non \\
	&=&   \frac{\ri}{2}\left(F^{\a\b}\right)^\rT \f F_{\a\b} + {\rm c.c.}~,
	\label{1.10}
\eea
where we have also introduced the real matrices $\X$ and $\U$ defined by
\bea
\f = \U + \ri \,\X~,\label{1.11}
\eea
with $\X$ being positive definite.
The model described by \eqref{1.10} has two fundamental properties: (i) its duality group is  $\sSp(2n, {\mathbb R})$ (see, e.g. \cite{KT2} for the technical details); and (ii) it is Weyl invariant. The induced action must respect these properties.

This paper is organised as follows. In section \ref{section2} we compute the
logarithmically divergent part of the effective action obtained by integrating out the vector fields in the model \eqref{1.10}. Generalisations of our analysis and open problems are briefly discussed in section \ref{section3}. The main body of the paper is accompanied by three
technical appendices. In appendix \ref{appendixA} we collect necessary facts about the Hermitian symmetric space $\sSp(2n, {\mathbb R} )/ \sU(n)$.
Appendix \ref{appsect:AlternativeFieldRedefinition} provides an alternative calculation of the induced action compared with that given in subsection \ref{sct:HeatKernelTechniques}. Appendix
\ref{appsect:CurvedSpaceBasisStructures} provides a complete list of the structures introduced in (\ref{aln:tra2(x,x)M,NResult}).

%%%%%%%%%%%%%%%%%%%%%%%%%%%%%%%
%%%%%%%%%%%%%%%%%%%%%%%%%%%%%

\section{Computing the induced action}\label{section2}

In this section we compute the logarithmically divergent part of the effective action,
$\G [\f, \bar \f]$, defined by
\begin{align}
{\rm e}^{\ri \G [\f, \bar \f]} &=\int [{\mathfrak D} A] \,
\d \Big( \eta - \c (A) \Big) \, {\rm Det} \,(\D_{\rm gh})\,
\re^{\ri S[A; \f, \bar \f] }~.
\label{EE}
 \end{align}
 Here $S[A;\f,\fb]$ is the classical action corresponding to (\ref{1.10}),
 \bea
	S[A;\f,\fb]&=&  \inte^4x\,\sqrt{-g} \,L^{(n)}(F;\f,\fb)~,
\eea
 $\c(A)$ denotes a gauge fixing condition,
 $\D_{\rm gh}$ the corresponding Faddeev-Popov operator \cite{FP},
 and $\eta$ an arbitrary background field. Since the effective action is independent of $\eta$,  this field can be integrated out with some weight that we choose to be
 \bea
 {\rm exp} \Big(-\frac{\ri}{2} \inte^4x\,\sqrt{-g}\,\eta^\rT \X \eta \Big)~.
 \label{weight}
\eea

In general, the logarithmically divergent part of the effective action has the form
\bea
\G_{\infty} = - \frac{\ln \L}{(4\pi)^2} \inte^4x\,\sqrt{-g}\, (a_2)_{\rm total} ~,
\eea
 where $ (a_2)_{\rm total} $ denotes the appropriate sum of diagonal DeWitt coefficients.
 We identify the induced action with $ \inte^4x\,\sqrt{-g}\, (a_2)_{\rm total} $,
 modulo an overall numerical coefficient.

%%%%%%%%%%%%%%%%%%%%%%%%%%%%%%
%%%%%%%%%%%%%%%%%%%%%%%%%%%%%%%

\subsection{Quantisation}\label{sct:Quantisation}

We choose the simplest gauge-fixing condition
\begin{equation}
	\c(A)=\nab^mA_m~,
	\end{equation}
which leads to the ghost operator
\begin{equation}
	\D_{\rm gh}:=\Box\id~,\label{eqn:GhostOperator}
\end{equation}
with $\id$ the $n\times n$ unit matrix.
Integrating the right-hand side of \eqref{EE}
with the weight functional \eqref{weight}
leads to the gauge-fixing term
\bea
S_{\text{G.F.}}[A;\f,\fb]=-\hf \inte^4x\,\sqrt{-g}\,(\nab^mA_m)^\rT \X(\nab^nA_n)~.
\eea
As a result, the gauge-fixed action becomes
\begin{equation}
	S_{\text{quadratic}}[A;\f,\fb]=\frac{1}{2}\inte^4x\,\sqrt{-g}\,A_{\hm}\D^{\hm\hn}A_{\hn}~,\label{eqn:QuadraticOperatorDefn}
\end{equation}
where here we have introduced hatted indices corresponding to a pair of spacetime and internal indices $A_{\hm}:=(A_{mi})$, $\D^{\hm\hn}:=(\D^{mi,nj})$. Contractions over hatted indices encode summations over both indices, however the position of the hatted indices (up or down) indicates only the position of the spacetime indices, internal indices are always understood as matrix multiplication. The non-minimal operator $\D^{\hm\hn}$ is defined as:
\begin{subequations} \label{N-MO}
\begin{align}
	\D^{mi,nj}&:=\X^{ij}g^{mn}\Box+V^{mi,p,nj}\nab_p-R^{mn}\X^{ij}~,
	\qquad \Box:=\nab^m\nab_m~,
	\label{aln:Non-MinOperatorDefn}\\
	V^{mi,p,nj}&:=(\nab^p\X^{ij})g^{mn}-(\nab^n\X^{ij})g^{mp}+(\nab^m\X^{ij})g^{pn}-(\nab_q\U^{ij})\ve^{mpnq}\label{aln:Non-MinOperatorDefnV}~.
\end{align}
\end{subequations}
From here onward matrix indices will be suppressed, unless there may be ambiguity or confusion.
The one-loop effective action is specified by
\begin{equation}
	\G^{(1)}[\f,\fb]=\frac{\ri}{2} \Tr\ln{\D} -\ri\Tr\ln\D_{\rm gh}~.\label{eqn:NonMinOneLoopEffectiveAction}
\end{equation}
Since $\X$ is symmetric and positive definite, due to (\ref{eqn:DilatonAxionConstraints}), its inverse $\X^{-1}$, square root $\X^{1/2}$ and inverse square root $\X^{-1/2}$ are well-defined. We perform a local field redefinition in
the path integral:
\begin{equation}\label{eqn:ConventionalFieldRedefinition}
	A_{m} \rightarrow \X^{-1/2}A_{m}~,
	\end{equation}
so that the operator which appeared in (\ref{eqn:QuadraticOperatorDefn})
becomes\footnote{In appendix \ref{appsect:AlternativeFieldRedefinition}, we provide the results for an alternative field redefinition which leads to an equivalent logarithmic divergence up to total derivative.}
\begin{equation}
	\~{\D}^{\hm\hn}=\X^{-1/2}\D^{mn}\X^{-1/2}~.
\end{equation}
Inserting the explicit form of $\D^{\hm\hn}$ from (\ref{aln:Non-MinOperatorDefn}) and (\ref{aln:Non-MinOperatorDefnV}), the $\~{\D}^{\hm}_{\ph{\hm}\hn}$ operator is now minimal:
\begin{subequations}
\begin{align}
	\~{\D}^{\hm}_{\ph{\hm}\hn}&=\id\,\d^m_{\ph{m}n}\Box+Q^{\hm}_{\ph{\hm}p\hn}\nab^p+T^{\hm}_{\ph{\hm}\hn}~,\label{aln:MinimalOperatorForm}\\
	Q^{\hm}_{\ph{\hm}p\hn}&:=-2\left(\nab_p\X^{1/2}\right)\X^{-1/2}\d^m_{\ph{m}n}+\X^{-1/2}\,V^m_{\ph{m}pn}\,\X^{-1/2}~,\label{aln:MinimalOperatorQDefn}\\
	T^{\hm}_{\ph{\hm}\hn}&:=-\left(\Box \X^{1/2}\right)\d^m_{\ph{m}n} +2\left(\nab^p\X^{1/2}\right)\X^{-1/2}\left(\nab_p\X^{1/2}\right)\X^{-1/2}\d^m_{\ph{m}n}\non\\
	&\quad-\X^{-1/2}\,V^m_{\ph{m}pn}\X^{-1/2}\left(\nab^p\X^{1/2}\right)\X^{-1/2}-R^m_{\ph{m}n}\id~.\label{aln:MinimalOperatorTDefn}
\end{align}
\end{subequations}
After our field redefinition the one-loop effective action is given by
\begin{equation}
	\G^{(1)}[\f,\fb]=\frac{\ri}{2}\Tr\ln\~{\D}-\ri\Tr\ln\D_{\rm gh}~.\label{eqn:OneLoopEffectiveAction}
\end{equation}

%%%%%%%%%%%%%%%%%%%%%%%%%%%%%%%%
%%%%%%%%%%%%%%%%%%%%%%%%%%%%%%%%%

\subsection{Heat kernel calculations}\label{sct:HeatKernelTechniques}

Since the operator $\tilde{\D}^{\hm}_{\ph{\hm}\hn}$ defined by
(\ref{aln:MinimalOperatorForm}) is minimal,
 we can proceed with the standard heat kernel technique in curved space, by bringing it to the form:
\begin{subequations}
\begin{align}
	\~{\D}^{\hm}_{\ph{\hm}\hn}&=\big(\hat{\nab}^p\hat{\nab}_p\big)^{\hm}_{\ph{\hm}\hn}+\hat{P}^{\hm}_{\ph{\hm}\hn}~,\label{aln:SchwingerDeWittOperatorForm}\\
	\hat{P}^{\hm}_{\ph{\hm}\hn}&:=-\frac{1}{2}\nab^pQ^{\hm}_{\ph{\hm}p\hn}-\frac{1}{4}\,Q^{\hm}_{\ph{\hm}q{\hp}}Q^{\hp q}_{\ph{\hp q}\hn}+T^{\hm}_{\ph{\hm}\hn}~.
\end{align}
\end{subequations}
The generalised covariant derivative $\hat{\nab}_m$ introduced above is defined to act on a column matrix $A^{\hm} =(A^m_{\ph{m}i})$ as
\begin{equation}
	\big(\hat{\nab}_pA\big)^{\hm}:=\nab_pA^{\hm}+\frac{1}{2}\,Q^{\hm}_{\ph{m}p\hn}A^{\hn}~.
\end{equation}
The generalised covariant derivatives have no torsion, meaning
\begin{equation}
		\big[\hat{\nab}_p,\hat{\nab}_q\big]
		A^{\hm}=\hat{R}^{\hm}_{\ph{\hm}\hn pq}A^{\hn}~,
\end{equation}
with $\hat{R}^{\hm}_{\ph{\hm}\hn pq}$ some generalised curvature anti-symmetric in $p,q$. Explicitly it has the form
\begin{equation}
	\hat{R}^{\hm}_{\ph{\m}\hn pq}=R^m_{\ph{m}npq}\id+\frac{1}{2}\nab_pQ^{\hm}_{\ph{\hm}q\hn}-\frac{1}{2}\nab_qQ^{\hm}_{\ph{\hm}p\hn}+\frac{1}{4}\,Q^{\hm}_{\ph{\hm}p\hr}Q^{\hr}_{\ph{\hr}q\hn}-\frac{1}{4}\,Q^{\hm}_{\ph{\hm}q\hr}Q^{\hr}_{\ph{\hr}p\hn}~.\label{eqn:GeneralisedCurvatureExplicit}
\end{equation}
Using the standard Schwinger-DeWitt formalism
\cite{DeWitt:1964mxt, BV, Avramidi, DeWitt:2003pm, Vassilevich}
in curved spacetime for an operator of the form (\ref{aln:SchwingerDeWittOperatorForm}), in the coincidence limit the DeWitt coefficient traced over matrix indices,
$(a_2)^{\~\D}(x,x)$, is given by
\begin{align}
	(a_2)^{\~\D}(x,x)=&\bigg(\frac{1}{45}R^{mnpq}R_{mnpq}-\frac{1}{45}R^{mn}R_{mn}+\frac{1}{18}R^2+\frac{2}{15}\Box R\bigg)\tr\id\non\\
	&+\frac{1}{12}\hat{R}^{\hm \hn pq}\hat{R}_{\hn \hm pq}+\frac{1}{6}\big(\hat{\nab}^p\hat{\nab}_p\hat{P}\big)^{\hm}_{\ph{\hm}\hm}+\frac{1}{2}\big(\hat{P}^2\big)^{\hm}_{\ph{\hm}\hm}+\frac{1}{6}R\hat{P}^{\hm}_{\ph{\hm}\hm}~,\label{aln:SchwingerDeWitta2}
\end{align}
where `$\tr$' denotes the matrix trace. Similarly for the ghost operator (\ref{eqn:GhostOperator}), the corresponding traced
DeWitt coefficient $(a_2)^{\D_{gh}}(x,x)$ (noting that the generalised curvature vanishes) is
\begin{equation}
	(a_2)^{\D_{gh}}(x,x)=\bigg(\frac{1}{180}R^{mnpq}R_{mnpq}-\frac{1}{180}R^{mn}R_{mn}+\frac{1}{72}R^2+\frac{1}{30}\Box R\bigg)\tr\id~,\label{eqn:SchwingerDeWitta2ghost}
\end{equation}
which contains purely gravitational components. Armed with the set of equations (\ref{aln:SchwingerDeWittOperatorForm} -- \ref{eqn:GeneralisedCurvatureExplicit}), we expand out $(a_2)^{\~\D}(x,x)$ (\ref{aln:SchwingerDeWitta2}) explicitly in terms of $Q^{\hm}_{\ph{\hm}p\hn}$ (\ref{aln:MinimalOperatorQDefn}) and $T^{\hm}_{\ph{\hm}\hn}$ (\ref{aln:MinimalOperatorTDefn})
\begin{align}
	(a_2)^{\~\D}(x,x)=&\bigg(-\frac{11}{180}R^{mnpq}R_{mnpq}-\frac{1}{45}R^{mn}R_{mn}+\frac{1}{18}R^2+\frac{2}{15}\Box R\bigg)\tr\id\non\\
	&+\frac{1}{6}R^{p\ph{m}qr}_{\ph{p}m}\left(\nab_qQ^{mi}_{\ph{mi}r,pi}\right)+\frac{1}{12}R^{p\ph{m}qr}_{\ph{p}m}Q^{mi}_{\ph{mi}q\hs}Q^{\hs}_{\ph{\hs}r,pi}-\frac{1}{12}R\left(\nab^pQ^{\hm}_{\ph{\hm}p\hm}\right)\non\\
	&-\frac{1}{24}R\,Q^{\hm}_{\ph{\hm}q\hp}Q^{\hp q}_{\ph{\hp q}\hm}+\frac{1}{6}R\,T^{\hm}_{\ph{\hm}\hm}-\frac{1}{12}\Box\nab^pQ^{\hm}_{\ph{\hm}p\hm}-\frac{1}{12}\left(\Box Q^{\hm}_{\ph{\hm}q\hp}\right)Q^{\hp q}_{\ph{\hp q}\hm}\non\\
	&-\frac{1}{24}\left(\nab_qQ^{\hm}_{\ph{\hm}r\hp}\right)\big(\nab^qQ^{\hp r}_{\ph{\hp r}\hm}\big)-\frac{1}{24}\left(\nab^qQ^{\hm}_{\ph{\hm}r\hp}\right)\big(\nab^rQ^{\hp}_{\ph{\hp}q\hm}\big)+\frac{1}{8}\left(\nab^rQ^{\hm}_{\ph{\hm}r\hp}\right)\big(\nab^sQ^{\hp}_{\ph{\hp}s\hm}\big)\non\\
	&+\frac{1}{24}\left(\nab^qQ^{\hm}_{\ph{\hm}r\hp}\right)Q^{\hp}_{\ph{\hp}q\hs}Q^{\hs r}_{\ph{\hs r}\hm}-\frac{1}{24}\left(\nab^qQ^{\hm}_{\ph{\hm}r\hp}\right)Q^{\hp r}_{\ph{\hp r}\hs}Q^{\hs}_{\ph{\hs}q\hm}\non\\
	&+\frac{1}{8}\left(\nab^rQ^{\hm}_{\ph{\hm}r\hp}\right)Q^{\hp}_{\ph{\hp}t\hs}Q^{\hs t}_{\ph{\hs t}\hm}+\frac{1}{96}\,Q^{\hm}_{\ph{\hm}q\hs}Q^{\hs}_{\ph{\hs}r\hp}Q^{\hp q}_{\ph{\hp q}{\hat t}}Q^{{\hat t}r}_{\ph{{\hat t}r}\hm}+\frac{1}{48}\,Q^{\hm}_{\ph{\hm}q\hp}Q^{\hp q}_{\ph{\hp q}\hr}Q^{\hr}_{\ph{\hr}s{\hat t}}Q^{{\hat t}s}_{\ph{{\hat t}s}\hm}\non\\
	&-\frac{1}{2}\left(\nab^pQ^{\hm}_{\ph{\hm}p\hq}\right)T^{\hq}_{\ph{\hq}\hm}-\frac{1}{4}T^{\hm}_{\ph{\hm}\hr}Q^{\hr}_{\ph{\hr}p\hq}Q^{\hq p}_{\ph{\hq p}\hm}+\frac{1}{6}\Box T^{\hm}_{\ph{\hm}\hm}+\frac{1}{2}(T^2)^{\hm}_{\ph{\hm}\hm}~.
\end{align}
Using the definitions of of $Q^{\hm}_{\ph{\hm}p\hn}$ (\ref{aln:MinimalOperatorQDefn}) and $T^{\hm}_{\ph{\hm}\hn}$ (\ref{aln:MinimalOperatorTDefn}), we perform the laborious task of expanding $(a_2)^{\~\D}(x,x)$ in terms of the matrices $\X$ and $\U$. It reduces to the following form:
\begin{align}
	(a_2)^{\~\D}(x,x)=\tr\bigg[&\frac{7}{24}T_1+\frac{1}{48}T_2-\frac{5}{12}T_3+\frac{1}{12}T_4+\frac{7}{12}T_5+\frac{5}{24}T_6-\frac{1}{24}T_7+\frac{5}{24}T_8+\frac{7}{24}T_{9}\non\\
	&+\frac{1}{48}T_{10}+\frac{1}{4}T_{11}+\frac{1}{4}T_{12}-\frac{1}{2}T_{13}+\frac{1}{2}T_{14}-\frac{1}{2}T_{15}-\frac{1}{2}T_{16}-\frac{1}{2}T_{17}-\frac{1}{2}T_{18}\non\\
	&+\frac{1}{6}T_{19}+\frac{1}{6}T_{20}+\cX\id+\nab_m\cY^m+\Box \cZ\bigg]~,\label{aln:tra2(x,x)M,NResult}
\end{align}
where the contributions $T_1 , \dots , T_{20}$
are listed in appendix \ref{appsect:CurvedSpaceBasisStructures}. We have also introduced:
\begin{subequations}
\begin{align}
	\cX:=&-\frac{11}{180}R^{mnpq}R_{mnpq}+\frac{43}{90}R^{mn}R_{mn}-\frac{1}{9}R^2-\frac{1}{30}\Box R~,\\
	\cY^m:=&-\frac{1}{4}\X^{-1}(\nab^m\X)\X^{-1}(\nab^n\X)\X^{-1}(\nab_n\X)+\frac{1}{4}\X^{-1}(\nab^m\X)\X^{-1}(\nab^n\U)\X^{-1}(\nab_n\U)\non\\
	&+\frac{1}{12}\X^{-1}(\nab^n\X)\X^{-1}(\nab^m\U)\X^{-1}(\nab_n\U)+\frac{1}{12}\X^{-1}(\nab^n\X)\X^{-1}(\nab_n\U)\X^{-1}(\nab^m\U)\non\\
	&+\frac{1}{6}\X^{-1}(\nab^m\X)\X^{-1}(\Box \X)-\frac{1}{6}\X^{-1}(\nab^m\U)\X^{-1}(\Box \U)\non\\
	&-\frac{1}{3}\Big(R^{mn}-\frac{1}{2}Rg^{mn}\Big)\X^{-1}(\nab_n\X)~,\label{aln:NabTotalDerivativeContribution}\\
	\cZ:=&-\frac{1}{6}\X^{-1}(\nab^m\U)\X^{-1}(\nab_m\U)-\frac{1}{3}\X^{-1}(\Box \X)+\frac{4}{3}\X^{-1/2}(\Box \X^{1/2})\non\\
	&+\frac{4}{3}\X^{-1}(\nab_n\X)\X^{-1/2}(\nab^n\X^{1/2})~.\label{aln:BoxTotalDerivativeContribution}
\end{align}
\end{subequations}
The total DeWitt coefficient corresponding to the logarithmic divergence of the effective action (\ref{eqn:OneLoopEffectiveAction}) is given by
\begin{equation}
	(a_2)_{\rm total}=(a_2)^{\~\D}(x,x)-2(a_2)^{\D_{gh}}(x,x)~,
\end{equation}
where $(a_2)^{\D_{gh}}(x,x)$ was given in  (\ref{eqn:SchwingerDeWitta2ghost}). Recalling the expression for $\X$ and $\U$ in terms of the original fields $\f$ and its conjugate $\fb$ (\ref{1.11}), and defining
\begin{align}
	\cD^2\f&:=\Box\f+\ri (\nab^m\f)\X^{-1}(\nab_m\f)~,\qquad \cD^2\fb:=\Box\fb-\ri (\nab^m\fb)\X^{-1}(\nab_m\fb)~,
\end{align}
the total  DeWitt coefficient
is given by
\begin{align}
	(a_2)_{\rm total}&=n\bigg(\frac{1}{10}F-\frac{31}{180}G-\frac{1}{10}\Box R\bigg)
		\non\\
	&+\frac{1}{4}\tr\Big[\X^{-1}(\cD^2\f) \X^{-1}(\cD^2\fb)-2\Big(R^{mn}-\frac{1}{3}Rg^{mn}\Big)\X^{-1}(\nab_m\f)\X^{-1}(\nab_n\fb)\Big]\non\\
	&+\frac{1}{24}\tr\Big[\X^{-1}(\nab^m\f)\X^{-1}(\nab_m\fb)\X^{-1}(\nab^n\f)\X^{-1}(\nab_n\fb)\Big]\non\\
	&+\frac{1}{48}\tr\Big[\X^{-1}(\nab^m\f)\X^{-1}(\nab^n\fb)\X^{-1}(\nab_m\f)\X^{-1}(\nab_n\fb)\Big]~,\label{aln:tra2(x,x)FinalResult}
\end{align}
where $F$ is the square of the Weyl tensor, $G$ is the Euler density,
\bea
F=R^{mnpq}R_{mnpq}-2R^{mn}R_{mn}+\frac{1}{3}R^2~,
\quad G=R^{mnpq}R_{mnpq}-4R^{mn}R_{mn}+R^2~,
\eea
and we have removed the total derivative pieces $\tr\big[\nab_m\cY^m\big]$ and $\tr\big[\Box \cZ\big]$ since they do not contribute to the induced action $\inte^4x\,\sqrt{-g}\,(a_2)_{\rm total}$.
The $\Box R$ in \eqref{aln:tra2(x,x)FinalResult} is also a total derivative and can be omitted.

Setting $n=1$
in \eqref{aln:tra2(x,x)FinalResult}
yields the expected result
derived
in \cite{BPT,Osborn}
\begin{align}
	(a_2)_{\rm total}&=\frac{1}{10}F-\frac{31}{180}G-\frac{1}{10}\Box R \non\\
	&+\frac{1}{4(\rm{Im}\f)^2}\Big[\cD^2\f \cD^2\fb-2\Big(R^{mn}-\frac{1}{3}Rg^{mn}\Big)\nab_m\f\nab_n\fb\Big]\non\\
	&+\frac{1}{48(\rm{Im}\f)^4}\Big[
	\nab^m\f\nab_m\f\nab^n\fb\nab_n\fb
	+2\nab^m\f\nab_m\fb\nab^n\f\nab_n\fb\Big]~.
\end{align}

%%%%%%%%%%%%%%%%%%%%%%%%%
%%%%%%%%%%%%%%%%%%%%%%%%%

\subsection{Geometric expression for the induced action}

To recast
(\ref{aln:tra2(x,x)FinalResult}) in terms of geometric
objects defined on the Hermitian symmetric space
$\sSp(2n,\dsR)/\sU(n)$,
here we analyse the dependence of
\eqref{aln:tra2(x,x)FinalResult} on the symmetric matrix
$\f=\left(\f^{ij}\right)=\left(\f^{ji}\right) \equiv (\f^I)$ and its conjugate $\fb=\left(\fb^{\,\ib\jb}\right)=\left(\fb^{\,\jb\ib}\right) \equiv (\bar \f^{\bar I})$.
%The reader is referred, e.g., to \cite{WB,Buchbinder:1998qv} for a pedagogical review of K\"ahler geometry.

 We make the standard choice for
  K\"ahler potential  ${\mathfrak K}(\f^{ij},\fb^{\,\ib\jb})$
  on $\sSp(2n,\dsR)/\sU(n)$
\begin{equation}
	{\mathfrak K}(\f,\fb):=-4\tr\ln\X~,
\end{equation}
which is well defined since $\X$ is a positive definite matrix.
The group  $\sSp(2n,\dsR)$ acts on  $\sSp(2n,\dsR)/\sU(n)$ by fractional linear transformations \eqref{A.18}. Given such a transformation, the K\"ahler potential
changes as
\begin{equation}
	{\mathfrak K}(\f,\fb)\rightarrow {\mathfrak K}(\f,\fb)+\L(\f)+\bar{\L}(\fb)~,
\end{equation}
in accordance with \eqref{aln:XiTransformationLaw}. Therefore, the K\"ahler metric is invariant under arbitrary  $\sSp(2n,\dsR)$ transformations.

The  K\"ahler metric is given by\footnote{The partial derivatives with respect to symmetric matrices $\f =(\f^{ij}) $
and $\fb=(\fb^{\,\ib\jb})$
are defined by $\rd {\mathfrak K}(\f,\fb) = \rd \f^{ij}
\frac{\pa {\mathfrak K}(\f,\fb)}{\pa \f^{ij} } + \rd \bar \f^{\bar i \bar j}
\frac{\pa {\mathfrak K}(\f,\fb)}{\pa \f^{\bar i \bar j} }$, and therefore
$\frac{\pa \f^{kl}}{\pa \f^{ij}} = \d^k_{(i} \d^l_{j)} $. Symmetrisation of $n$ indices includes a $1/n!$ factor. Vertical bars are notation to exclude indices contained between them from a separate symmetrisation, for example, $(i_1|(i_2i_3)|i_4)$.}
\begin{equation}
	{\mathfrak g}_{ij,\kb\lb}={\mathfrak g}_{(ij),(\kb\lb)}=\frac{\pd^2 {\mathfrak K}}{\pd \f^{ij}\pd \fb^{\,\kb\lb}}=(\X^{-1})_{i (\kb}(\X^{-1})_{\lb)j}~,\label{eqn:KahlerMetric}
\end{equation}
where $(i_1\cdots i_n)$ denotes symmetrisation in indices $i_1,\ldots,i_n$.
Note that pairs of indices are symmetrised over due to $\f$ being symmetric (\ref{eqn:DilatonAxionConstraints}). Here and in what follows, we use the notation
\begin{equation}
	{\mathfrak K}_{i_1i_2,\ldots,i_{2p-1}i_{2p},\ib_1\ib_2,\ldots,\ib_{2q-1}\ib_{2q}}=\frac{\pd^{p+q} {\mathfrak K}}{\pd \f^{i_1i_2}\cdots\pd \f^{i_{2p-1}i_{2p}}\pd \fb^{\,\ib_1\ib_2}\cdots\pd \fb^{\,\ib_{2q-1}\ib_{2q}}}~.
\end{equation}
In accordance with \eqref{18},
the Christoffel symbols are given by
\begin{equation}
	\G^{i_1i_2}_{\ph{i_1i_2}i_3i_4,i_5i_6}={\mathfrak g}^{i_1i_2,\ib_7\ib_8}{\mathfrak K}_{i_3i_4,i_5i_6,\ib_7\ib_8}~,\qquad\G^{\ib_1\ib_2}_{\ph{\ib_1\ib_2}\ib_3\ib_4,\ib_5\ib_6}=(\G^{i_1i_2}_{\ph{i_1i_2}i_3i_4,i_5i_6})^*~,
\end{equation}
and the Riemann curvature tensor is
\begin{equation}
	{\mathfrak R}_{i_1i_2,\ib_3\ib_4,i_5i_6,\ib_7\ib_8}={\mathfrak K}_{i_1i_2,i_5i_6,\ib_3\ib_4,\ib_7\ib_8}-{\mathfrak g}^{i_9i_{10},\ib_{11}\ib_{12}}{\mathfrak K}_{i_1i_2,i_5i_6,\ib_{11}\ib_{12}}{\mathfrak K}_{i_9i_{10},\ib_3\ib_4,\ib_7\ib_8}~.
\end{equation}
Noting that the inverse K\"ahler metric of (\ref{eqn:KahlerMetric}) is
\begin{equation}
	{\mathfrak g}^{ij,\kb\lb}=\X^{i(\kb}\,\X^{\lb) j}~,
\end{equation}
one can calculate the Christoffel symbols, Riemann curvature tensor and Ricci tensor for the metric considered in (\ref{eqn:KahlerMetric}) and we find:
\begin{subequations}
\begin{align}
	\G^{i_1i_2}_{\ph{i_1i_2}i_3i_4,i_5i_6}&=\ri \d^{(i_1}_{\ph{(i_1}(i_3}(\X^{-1})_{i_4)(i_5}\d^{i_2)}_{\ph{i_2)}i_6)}~,\label{aln:KahlerChristoffel}\\
	\G^{\ib_1\ib_2}_{\ph{\ib_1\ib_2}\ib_3\ib_4,\ib_5\ib_6}&=-\ri \d^{(\ib_1}_{\ph{(\ib_1}(\ib_3}(\X^{-1})_{\ib_4)(\ib_5}\d^{\ib_2)}_{\ph{\ib_2)}\ib_6)}~,\\
	{\mathfrak R}_{i_1i_2,\ib_3\ib_4,i_5i_6,\ib_7\ib_8}&=\frac{1}{2}(\X^{-1})_{(\ib_8|(i_1}(\X^{-1})_{i_2)(\ib_3}(\X^{-1})_{\ib_4)(i_5}(\X^{-1})_{i_6)|\ib_7)}~,\label{aln:KahlerCurvature}\\
	{\mathfrak R}_{i_1i_2,\ib_3\ib_4}&=-\frac{n}{2}(\X^{-1})_{i_1(\ib_3}(\X^{-1})_{\ib_4)i_2}=-\frac{n}{2}{\mathfrak g}_{i_1i_2,\ib_3\ib_4}~.
\end{align}
\end{subequations}
The latter relation means that $\sSp(2n,\dsR)/\sU(n)$ is an Einstein space.

As pointed out at the beginning of this subsection, the complex variables $\f$  and
their conjugates $\bar \f$ can be
viewed either as symmetric matrices $\f =(\f^{ij})$  and $\bar \f =( \fb^{\ib\jb})$
or as vector columns
$\f = (\f^I)$ and $\bar \f =( \fb^{\Ib})$,  with
$I,\Ib=1,\ldots,\hf {n(n+1)}$.
 Resorting to the latter notation,  the geometric structures (\ref{eqn:KahlerMetric}) and
(\ref{aln:KahlerChristoffel} -- \ref{aln:KahlerCurvature}) can be used to recast
(\ref{aln:tra2(x,x)FinalResult}) in the form:
\begin{subequations}\label{aln:tra2(x,x)KahlerResult}
\begin{align}
		(a_2)_{\rm total}&=n\bigg(\frac{1}{10}F-\frac{31}{180}G-\frac{1}{10}\Box R\bigg)
				\non\\
		&+\frac{1}{4}{\mathfrak g}_{I\Jb}(\f,\fb)\Big[\cD^2\f^{I}\cD^2\fb^{\Jb}-2\Big(R^{mn}-\frac{1}{3}Rg^{mn}\Big)\nab_m\f^{I}\nab_n\fb^{\Jb}\Big]
		\\
		&+\frac{1}{24}{\mathfrak R}_{I\Jb K \Lb}(\f,\fb)\Big[2\nab^m\f^{I}\nab_m\fb^{\Jb}\nab^n\f^{K}\nab_n\fb^{\Lb}+\nab^m\f^{I}\nab^n\fb^{\Jb}\nab_m\f^{K}\nab_n\fb^{\Lb}\Big]~,\non	
\end{align}
where
\begin{align}
	\cD^2\f^I=\Box\f^I+\G^I_{\ph{I}JK}\nab^m\f^J\nab_m\f^K~,\qquad \cD^2\fb^{\Ib}=\Box\fb^{\Ib}+\G^{\Ib}_{\ph{\Ib}\Jb\Kb}\nab^m\fb^{\Jb}\nab_m\fb^{\Kb}~.
\end{align}
\end{subequations}
Every isometry transformation \eqref{A.18} acts on $\nabla \f$ and $\cD^2 \f$ as follows:
\begin{subequations}\label{aln:CovariantTransformationLaw}
\begin{align}
	\nab^m\f'&=\big( (C\f+D)^{-1}\big)^{\rT}(\nab^m\f)(C\f+D)^{-1}~,\\
	\cD^2\f'&=\big( (C\f+D)^{-1}\big)^{\rT}(\cD^2\f)(C\f+D)^{-1}~.
\end{align}
\end{subequations}
It is now seen that the induced action defined by \eqref{aln:tra2(x,x)KahlerResult} is invariant under the isometry transformations   on $\sSp(2n,\dsR)/\sU(n)$.

%%%%%%%%%%%%%%%%%%%%%%%%%%%%%
%%%%%%%%%%%%%%%%%%%%%%%%%%%%%%

\section{Generalisations and open problems}\label{section3}

Relation \eqref{aln:tra2(x,x)KahlerResult}, which constitutes the induced action, is our
main result. The same structure also determines the Weyl anomaly of the effective action
\bea
\d_\s \G \propto
\frac{1}{(4\pi)^2} \inte^4x\,\sqrt{-g}\, \s \,(a_2)_{\rm total} ~.
\label{311}
\eea
It is well known that the purely gravitational part of this variation
satisfies the Wess-Zumino consistency condition \cite{Wess:1971yu}
\bea
[\d_{\s_2} , \d_{\s_1} ]\G =0~,
\label{WZ}
\eea
see, e.g., \cite{FT85,DeserS,Duff:1993wm} for a review.\footnote{The $\Box R$ term, which
contributes to  $(a_2)_{\rm total} $ in \eqref{311}, can be removed since it is generated
by a local counterterm $ \inte^4x\,\sqrt{-g}\, R^2$.}
The $\f$-dependent part of the Weyl anomaly will be discussed below.

As a generalisation of \eqref{1.4}, we can introduce a conformal higher-derivative $\s$-model associated with a Riemannian manifold
$(\cW^d, {\mathfrak g})$ parametrised by local coordinates $\vf^\m$. The action is
\begin{subequations}\label{3.1}
\begin{align}
	S=\inte^4x\,\sqrt{-g} \,&\bigg\{
		{\mathfrak g}_{\m \n }(\vf)\Big[\cD^2 \vf^{\m}\cD^2 \vf^{\n}
	-2\Big(R^{mn}-\frac{1}{3}Rg^{mn}\Big)\nab_m\vf^\m\nab_n\vf^{\n}\Big] \non \\
	&+ {\mathfrak F}_{\m\n \s \r} (\vf)	
\nab^m\vf^{\m} \nab_m\vf^{\n}  \nab^n\vf^{\s} \nab_n\vf^{\r}
\bigg\} ~,
\end{align}
where
\bea
\cD^2 \vf^\m := \Box \vf^\m + \G^\m_{\ph{\m}\n\s } \nabla^m \vf^\n \nabla_m \vf^\s ~, \qquad
\Box = \nabla^m \nabla_m~,
\eea
\end{subequations}
and $ {\mathfrak F}_{\m\n \s \r} (\vf)$ is a  tensor field of rank $(0,4)$ on $\cW^d$.
The Weyl invariance of the above action  follows from
\bea
\d_\s \bigg\{\sqrt{-g} \,
		{\mathfrak g}_{\m \n }(\vf)\Big[\cD^2 \vf^{\m}\cD^2 \vf^{\n}
	-2\Big(R^{mn}-\frac{1}{3}Rg^{mn}\Big)\nab_m\vf^\m\nab_n\vf^{\n}\Big]  \bigg\}\qquad\qquad \non \\
\qquad \qquad=4	\sqrt{-g} \nabla_m \bigg\{
{\mathfrak g}_{\m \n }(\vf) \Big[ \nabla_n \s \nabla^m\vf^\m \nabla^n \vf^\n
-\hf \nabla^m \s \nabla^n \vf^\m \nabla_n \vf^\n \Big] \bigg\}~.
\label{3.3}
\eea
In the case that the target space is K\"ahler, eq. \eqref{1.4}, the relation \eqref{3.3}
takes the form
\begin{align}
&\d_{\s}\bigg\{\sqrt{-g}\,{\mathfrak g}_{I\Jb}(\f,\fb)\Big[\cD^2\f^{I}\cD^2\fb^{\Jb}
-2\Big(R^{mn}-\frac{1}{3}Rg^{mn}\Big)\nab_m\f^{I}\nab_n\fb^{\Jb}\Big]\bigg\}\non\\
&=2\sqrt{-g}\,\nab_m\bigg\{ {\mathfrak g}_{I\Jb}(\f,\fb)\Big[\nab_n\s\Big(\nab^m\f^{I}\nab^n\fb^{\Jb}+\nab^n\f^{I}\nab^m\fb^{\Jb}\Big)-\nab^m\s\nab^n\f^{I}\nab_n\fb^{\Jb}\Big]\bigg\}~.
\end{align}

Choosing $\cW^d$ to be $\mathbb R$, specifying  ${\mathfrak g}_{\m \n }(\vf) $ and
${\mathfrak F}_{\m\n \s \r} (\vf)$ to be constant, respectively,
and restricting the background spacetime to be flat, the action \eqref{3.1} turns into
\bea
S = \int \rd^4x \, L~, \qquad L = (\pa^2 \vf)^2 +f (\pa^m \vf \pa_m \vf)^2 ~,
\eea
with $f$ a coupling constant, which is the model studied recently by Tseytlin \cite{Tseytlin:2022flu}.

Assuming the model \eqref{3.1} originates from an induced action of some theory,
the $\vf$-dependent part of the Weyl anomaly should have the form
\begin{align}
\d_\s \G \propto \inte^4x\,\sqrt{-g} \, & \s \bigg\{
		{\mathfrak g}_{\m \n }(\vf)\Big[\cD^2 \vf^{\m}\cD^2 \vf^{\n}
	-2\Big(R^{mn}-\frac{1}{3}Rg^{mn}\Big)\nab_m\vf^\m\nab_n\vf^{\n}\Big] \non \\
	& \qquad + {\mathfrak F}_{\m\n \s \r} (\vf)	
\nab^m\vf^{\m} \nab_m\vf^{\n}  \nab^n\vf^{\s} \nab_n\vf^{\r}
\bigg\} ~.
\end{align}
The anomaly satisfies the Wess-Zumino consistency condition \eqref{WZ}
as a consequence of the relation \eqref{3.3}.

It would be interesting to study renormalisation properties  of a higher-derivative theory in Minkowski space with Lagrangian of the form
\begin{align}
\cL (F, &\f , \bar \f)
=L^{(n)}(F;\f,\fb)
+f_1 {\mathfrak g}_{I\Jb}(\f,\fb)\cD^2\f^{I}\cD^2\fb^{\Jb} 	\non \\
		&+{\mathfrak R}_{I\Jb K \Lb}(\f,\fb)\Big[f_2 \pa^m\f^{I}\pa_m\fb^{\Jb}\pa^n\f^{K}\pa_n\fb^{\Lb}+f_3 \pa^m\f^{I}\pa^n\fb^{\Jb}\pa_m\f^{K}\pa_n\fb^{\Lb}\Big] +\dots~,
		\label{3.8}
\end{align}
where $L^{(n)}(F;\f,\fb) $ is given by \eqref{1.10}, the complex scalar fields $\f^I$ and their conjugates $\bar \f^{\bar I}$ parametrise $\sSp(2n, {\mathbb R} )/ \sU(n)$, and  $f_1, f_2 $ and $f_3$ are dimensionless coupling constants. All the structures in the $F$-independent part of \eqref{3.8} appear in the induced action \eqref{aln:tra2(x,x)KahlerResult}.
The ellipsis in \eqref{3.8} denotes other $\sSp(2n, {\mathbb R})$ terms that are
quartic in $\pa \f$ and $\pa \bar \f$, such as the K\"ahler metric squared  structure  in \eqref{F-structure}. Such terms are possible for $n>1$.
The renormalisation of the most general fourth-order sigma models with dimensionless couplings in four dimensions was studied in \cite{Buchbinder:1988ei, Buchbinder:1991jw}.
All couplings constants in \eqref{3.8} are dimensionless, and the freedom to choose them is
dictated by   $\sSp(2n, {\mathbb R})$. This implies that the theory with classical Lagrangian \eqref{3.8} is renormalisable at the quantum level.

It was discovered two years ago that Maxwell's theory possesses a one-parameter
conformal and $\sU(1)$ duality invariant deformation \cite{BLST,Kosyakov}; it was called
the ModMax theory in \cite{BLST}.
Using the methods developed in  \cite{GR2,GZ2,GZ3}, it can be
coupled to the dilaton-axion field \eqref{1.8} to result in a conformal and $\sSL(2,{\mathbb R})$ duality invariant model described by the Lagrangian \cite{K2021}
\begin{subequations} \label{3.9}
\bea
L_\g(F ; \f , \bar \f) =
- \hf \re^{-\vf}  \big( {\w} + {\bar \w}\big) (\cosh \g -1)
+  \re^{-\vf}  {\sqrt{\w\bar \w} }\,{\sinh \g }
+\frac{\ri}{2} \big(\f \,\w  - \bar \f \,\bar  \w \big)  ~,
\eea
where
\bea
\w=\a+\ri \b = F^{\a\b} F_{\a\b}~, \qquad
\a = \frac{1}{4} \, F^{ab} F_{ab}~, \qquad
\b = \frac{1}{4} \, F^{ab} \tilde{F}_{ab} ~,
\eea
\end{subequations}
and $\g$ is a non-negative coupling constant \cite{BLST}. For $\g= 0$ the
model \eqref{3.9} reduces to \eqref{1.8}.
A challenging problem is to compute an induced action generated by \eqref{3.9}.

\noindent
{\bf Acknowledgements:}\\
The work of SK is supported in part by the Australian Research Council, project No. DP200101944. The work of JP is supported by the Australian Government Research Training Program Scholarship.

%%%%%%%%%%%%%%%%%%%%%%%%%%%%%%%%%%
%%%%%%%%%%%%%%%%%%%%%%%%%%%%%%%%%

\appendix

\section{Hermitian symmetric space $\sSp(2n, {\mathbb R} )/ \sU(n)$} \label{appendixA}

In this appendix we collect necessary facts about the Hermitian symmetric space
$\sSp(2n, {\mathbb R} )/ \sU(n)$. Here the symplectic group is defined by
\bea
\sSp(2n, {\mathbb R} ) = \left\{ g \in \sGL(2n, {\mathbb R}) , \quad g^{\rm T} J g = J ~,
\quad J
=\left(
\begin{array}{cc}
0  & \quad {\mathbbm 1}_n\\
 -{\mathbbm 1}_n  &   \quad  0
\end{array}
\right)
 \right\}~.
 \label{A.1}
\eea
Its maximal compact subgroup, $H = \sSp(2n, {\mathbb R} ) \cap \sSO(2n)$, proves to be isomorphic to $\sU(n)$. One way to see this is to make use of the isomorphism
\bea
\sSp(2n, {\mathbb R} ) \cong \sSp(2n, {\mathbb C} ) \cap \sSU(n,n) \equiv \underline{G}~,
\label{A.2}
\eea
which is obtained by considering the bijective map
\bea
\vf: g ~\to ~\underline{g} = {\mathbb A}^{-1} g {\mathbb A}~,
\qquad
{\mathbb A} = \frac{1}{\sqrt{2}} \left(
\begin{array}{rc}
{\mathbbm 1}_n  ~& ~{\mathbbm 1}_n\\
 -\ri {\mathbbm 1}_n  ~& ~   \ri {\mathbbm 1}_n
\end{array}
\right) ~,
\label{A.3}
 \eea
for any $ g \in \sSp(2n, {\mathbb R} ) $. Each complex matrix $ \underline{g} = \vf (g) $ is characterised by the properties
\bea
 \underline{g}^{\rm T} J   \underline{g} = J~, \qquad
  \underline{g}^\dagger I_{n,n}  \underline{g} = I_{n,n} ~, \qquad
  I_{n,n}=
   \left(
\begin{array}{cc}
{\mathbbm 1}_n  ~& 0 \\
 0  ~& - {\mathbbm 1}_n
\end{array}
\right) ~,
\eea
and thus $\underline{g} \in  \sSp(2n, {\mathbb C} ) \cap \sSU(n,n) $. The inverse map $\vf^{-1}$ takes every group element $\underline{g}  \in  \sSp(2n, {\mathbb C} ) \cap \sSU(n,n) $ to some $g \in \sGL(2n, {\mathbb R}) $. For every element $h$ from the maximal compact subgroup, $h \in H = \sSp(2n, {\mathbb R} ) \cap \sSO(2n)$, its image
$\underline{h} = \vf (h) $ is unitary,
 $\underline{h}^\dagger \underline{h} ={\mathbbm 1}_{2n}$.
Simple calculations show that every $h \in H = \sSp(2n, {\mathbb R} ) \cap \sSO(2n)$
has the form
\bea
h =  \left(
\begin{array}{rc}
A  ~& \quad B \\
 -B  ~&  \quad A
\end{array}
\right) ~,\qquad A A^{\rm T} +B B^{\rm T } = {\mathbbm 1}_n~, \qquad
A B^{\rm T} = B A^{\rm T }~,
\eea
and for its image $\underline{h} =\vf (h)$ we obtain
\bea
\vf(h) = \left(
\begin{array}{cc}
A  -\ri B~& 0 \\
 0  ~& A +\ri B
\end{array}
\right) ~, \qquad (A +\ri B )(A +\ri B)^\dagger = {\mathbbm 1}_n~.
\label{A.6}
\eea

The group $\sSp(2n, {\mathbb R} )$ naturally acts on ${\mathbb C}^{2n}$. This action is extended to that on a complex Grassmannian ${\rm Gr}_{m, 2n} ({\mathbb C})$, with $0<m<2n$,
the space of $m$-planes through the origin in ${\mathbb C}^{2n}$.
Of special interest is the Grassmannian ${\rm Gr}_{n, 2n} ({\mathbb C})$. Given an $n$-plane $\cP \in {\rm Gr}_{n, 2n} ({\mathbb C})$, it can be identified with a $2n \times n$ matrix of rank $n$
\begin{subequations} \label{A.7}
\bea
\cP = \left(
\begin{array}{c}
 M \\
  N
  \end{array}
\right)~,
\eea
defined modulo equivalence transformations
\bea
\left(
\begin{array}{c}
 M \\
  N
  \end{array}
\right) \sim
\left(
\begin{array}{c}
 M R\\
  NR
  \end{array}
\right)~,\qquad R \in \sGL(n, {\mathbb C})~.
\eea
\end{subequations}
We denote by ${\mathfrak X} \subset {\rm Gr}_{n, 2n} ({\mathbb C})$ the collection of all $n$-planes satisfying the two conditions:
\begin{subequations} \label{A.8}
\bea
\cP^{\rm T} J \cP &=&0~; \label{A.8a}\\
\cP^\dagger (\ri J ) \cP &>& 0~. \label{A.8b}
\eea
\end{subequations}
Condition \eqref{A.8b} means that the Hermitian matrix $\cP^\dagger (\ri J ) \cP$ is positive definite. Condition \eqref{A.8a} means that $\cP$ is a Lagrangian subspace of
${\mathbb C}^{2n}$ with respect to the symplectic structure $J$.

By construction, the group $\sSp(2n, {\mathbb R} )$ naturally acts on ${\mathfrak X} $.
It turns out that this action is transitive, and $\mathfrak X$ can be identified with $\sSp(2n, {\mathbb R} )/ \sU(n)$. The simplest way to see this is to make use of the realisation $\underline G$ of  $\sSp(2n, {\mathbb R} )$, eq. \eqref{A.2}.  The picture changing
transformation \eqref{A.3} is accompanied with
\bea
\cP ~\to~ \underline{\cP}= {\mathbb A}^{-1} \cP
=\left(
\begin{array}{c}
 \underline{M} \\
  \underline{N}
  \end{array}
\right)~.
\label{A.9}
\eea
For the transformed $n$-plane $\underline{\cP}$ the conditions \eqref{A.8} take the form
\begin{subequations}\label{A.10}
\bea
\underline{\cP}^{\rm T} J \underline{\cP} &=&0~; \label{A.10a}\\
\underline{\cP}^\dagger I_{n,n}  \underline{\cP} &>& 0~. \label{A.10b}
\eea
\end{subequations}
The latter condition tells us that the matrix $\underline{M}$ in \eqref{A.9} is nonsingular,
and therefore
\bea
\left(
\begin{array}{c}
 \underline{M} \\
  \underline{N}
  \end{array}
\right) ~\sim~ \left(
\begin{array}{c}
{\mathbbm 1}_n \\
  \underline{N} \underline{M}^{-1}
  \end{array}
\right)~\equiv ~  \left(
\begin{array}{c}
{\mathbbm 1}_n \\
\psi
  \end{array}
\right)~.
\label{A.11}
\eea
In terms of the $n\times n$ matrix $\psi$,  the conditions \eqref{A.10} are equivalent to
\bea
\psi^{\rm T} = \psi~, \qquad {\mathbbm 1}_n > \psi^\dagger \psi ~.
\label{A.12}
\eea
The complex $n\times n$ matrix $\psi$ constrained by \eqref{A.12} and its conjugate $\bar \psi$ provide a global coordinate system for $\mathfrak X$.
Associated with $\psi$ and $\bar \psi$ is the group element
\bea
{\mathfrak S} (\psi, \bar \psi) =  \left(
\begin{array}{cc}
s~& \bar \psi \bar s  \\
 \psi s   ~& \bar s
\end{array}
\right) \in \sSp(2n, {\mathbb C} ) \cap \sSU(n,n) ~, \quad
s = \big( {\mathbbm 1}_n - \bar \psi \psi \big)^{-\hf} ~.
\eea
Its important property is
\bea
{\mathfrak S} (\psi, \bar \psi)
\cP_0
= \left(
\begin{array}{c}
s \\
\psi s
  \end{array}
\right) \sim
\left(
\begin{array}{c}
{\mathbbm 1}_n \\
\psi
  \end{array}
\right)~, \qquad \cP_0 := \left(
\begin{array}{c}
{\mathbbm 1}_n \\
0
\end{array}
\right)~.
\eea
Thus ${\mathfrak S} (\psi, \bar \psi) $ maps the ``origin''
$\cP_0 $ to the point \eqref{A.11} of $\mathfrak X$, and therefore
the group $\sSp(2n, {\mathbb C} ) \cap \sSU(n,n) $ acts transitively on
$\mathfrak X$. The isotropy subgroup of $\cP_0$ can be seen to consist of the group elements  \eqref{A.6}, which span $\sU(n)$. We conclude that
\bea
{\mathfrak X} = \frac{ \sSp(2n, {\mathbb C} ) \cap \sSU(n,n) }{\sU(n)}
=\frac{ \sSp(2n, {\mathbb R} ) }{\sU(n)}
~.
\label{A.15}
\eea

Making use of \eqref{A.9} -- \eqref{A.11}, we can reconstruct a generic element of $\mathfrak X$ in the original real realisation \eqref{A.1}.
\bea
\cP = {\mathbb A} \left(
\begin{array}{c}
{\mathbbm 1}_n \\
\psi
  \end{array}
\right) \sim
 \left(
\begin{array}{c}
{\mathbbm 1}_n +\psi \\
-\ri ({\mathbbm 1}_n -\psi )
  \end{array}
\right)~.
\eea
Since the matrices $ {\mathbbm 1}_n \pm \psi $ are non-singular, $\cP$ is equivalent to
\begin{subequations} \label{A.17}
\bea
\cP \sim
 \left(
\begin{array}{c}
\f \\
{\mathbbm 1}_n
  \end{array}
\right)~,
\qquad \f = \ri  \frac{
{\mathbbm 1}_n +\psi }
{ {\mathbbm 1}_n -\psi }~.
\eea
The properties of $\f$ follow from \eqref{A.8}:
\bea
\f^{\rm T} = \f~, \qquad \ri \big(\bar \f - \f\big) >0~.
\eea
\end{subequations}
In this paper we make use of the $\f$-parametrisation \eqref{A.17}
of the coset space \eqref{A.15}.

The group  $\sSp(2n,\dsR)$ acts on $\sSp(2n,\dsR)/\sU(n)$ by fractional linear transformations
\begin{equation}
	\f ~\to ~ \f'= (A\f+B)(C\f+D)^{-1}~,  \qquad
	\begin{pmatrix}
		A & \quad B\\
		C &  \quad D	
	\end{pmatrix}\in\sSp(2n,\dsR)~,
	\label{A.18}
\end{equation}
and therefore
\bea
	\rd \f'=\big( (C\f+D)^{-1}\big)^{\rT}\rd \f \,(C\f+D)^{-1}~.
	\label{A.18extra}
\eea
Using the definition of symplectic matrices, eq. \eqref{A.1}, one can show that
the positive-definite matrix $\X$, which is defined by \eqref{1.11}, transforms as follows:
 \begin{align}
	(\X')^{-1}&=(C\f+D)\X^{-1}(C\fb+D)^\rT=(C\fb+D)\X^{-1}(C\f+D)^\rT~.\label{aln:XiTransformationLaw}
\end{align}

Let $\cP_1$ and $\cP_2$ be two points in  $\sSp(2n,\dsR)/\sU(n)$. We associate with them the following two-point function:
\bea
{\mathfrak s}^2 (\cP_1, \cP_2) = - 4 {\rm Tr} \Big[ \big(\cP_1^\dagger J \bar \cP_2\big)
\big(\cP_2^{\rm T} J \bar \cP_2\big)^{-1} \big(\cP_2^{\rm T}  J \bar \cP_1\big)
\big(\cP_1^\dagger J \bar \cP_1\big)^{-1} \Big]~.
\label{A.20}
\eea
By construction, it is invariant under arbitrary $\sSp(2n,\dsR)$ transformations
\bea
\cP_{1,2}  ~\to ~ g \cP_{1,2} ~, \qquad g \in \sSp(2n,\dsR)~.
\eea
It is also invariant under arbitrary equivalence transformations \eqref{A.7},
\bea
\cP_{1,2}  ~\to ~ \cP_{1,2} R_{1,2}~,
\qquad R_{1,2} \in \sGL(n, {\mathbb C})~.
\label{A.22}
\eea
Therefore, ${\mathfrak s}^2 (\cP_1, \cP_2)$ is a  two-point function of
 $\sSp(2n,\dsR)/\sU(n)$ which is invariant under the isometry group  $\sSp(2n,\dsR)$.
Due to the invariance of ${\mathfrak s}^2 (\cP_1, \cP_2) $ under arbitrary right shifts
\eqref{A.22}, both $\cP_1$ and $\cP_2$ can be chosen to have the form \eqref{A.17}.
In the case that $\cP_1$ and $\cP_2$ are infinitesimally separated,
${\mathfrak s}^2 (\cP_1, \cP_2) $ becomes
\bea
\rd {\mathfrak s}^2 = {\rm Tr} \Big[ \rd \bar \f \,\X^{-1} \rd \f \,\X^{-1} \Big] ~,
\eea
which is the K\"ahler metric on  $\sSp(2n,\dsR)/\sU(n)$, eq. \eqref{eqn:KahlerMetric}.

%%%%%%%%%%%%%%%%%%%%%%%%%%%%%%%%%%%%%
%%%%%%%%%%%%%%%%%%%%%%%%%%%%%%%%%%%%%

\section{Alternative field redefinition}\label{appsect:AlternativeFieldRedefinition}

Here we describe an alternative calculation of $\Tr\ln{\D} $
 compared with that given in section \ref{sct:HeatKernelTechniques}.
Consider a path integral over two vector fields
\begin{align}
	{\rm Det} \,\D^{-1}
	=N\int[{\mathfrak D} B] [{\mathfrak D} A]\, \re^{\ri \inte^4x\,\sqrt{-g}\,B_{\hm}\D^{\hm\hn}A_{\hn}}
		\label{aln:AlternativePathIntegral}
\end{align}
with the  operator  $\D^{\hm\hn}$ given in
 (\ref{aln:Non-MinOperatorDefn}).
We perform an alternative  local field definition in the path integral
\begin{equation}
	A^{\hm} \rightarrow A^{\hm},\qquad B_{\hm} \rightarrow \X^{-1} B_m~,
\end{equation}
which was not possible for the original quadratic action (\ref{eqn:QuadraticOperatorDefn}). Now the operator which appeared in (\ref{aln:AlternativePathIntegral}) has the form
\begin{equation}
	\~{\D}^{\hm}_{\ph{m}\hn}:=\X^{-1}\D^m_{\ph{m}n}~,
\end{equation}
which, once expanded explicitly from (\ref{aln:Non-MinOperatorDefn}) and (\ref{aln:Non-MinOperatorDefnV}), is  minimal:
\begin{subequations}
	\begin{align}
		\~{\D}^{\hm}_{\ph{m}\hn}&=\id\,\d^m_{\ph{m}n}\Box+Q^m_{\ph{m}pn}\nab^p+T^m_{\ph{m}n}~,\\
		Q^{\hm}_{\ph{m}p\hn}&:=\X^{-1}\left(\nab_p\X\right)\d^m_{\ph{m}n}-\X^{-1}\left(\nab_n\X\right)\d^m_{\ph{m}p}+\X^{-1}\left(\nab^m\X\right)g_{pn}-\X^{-1}\left(\nab^q\U\right)\e^m_{\ph{m}pnq}~,\\
		T^{\hm}_{\ph{m}\hn}&:=-R^m_{\ph{m}n}\id~.
	\end{align}
\end{subequations}
Although both approaches of obtaining a minimal operator will lead to equivalent logarithmic divergences up to total derivative, the alternative field definition proves to be much easier to manage computationally, since it solely involves derivatives of $\X$, $\X^{-1}$ and $\U$, rather than $\X^{1/2}$ and $\X^{-1/2}$. Following the same procedure as section \ref{sct:HeatKernelTechniques}, the final result for $(a_2)_{\rm total}$
(including total derivative contributions)
 is
\begin{align}
	(a_2)_{\rm total}&=n\bigg(\frac{1}{10}F-\frac{31}{180}G-\frac{1}{10}\Box R\bigg)\non\\
	&+\frac{1}{4}\tr\Big[\X^{-1}(\cD^2\f) \X^{-1}(\cD^2\fb)-2\Big(R^{mn}-\frac{1}{3}Rg^{mn}\Big)\X^{-1}(\nab_m\f)\X^{-1}(\nab_n\fb)\Big]\non\\
	&+\frac{1}{24}\tr\Big[\X^{-1}(\nab^m\f)\X^{-1}(\nab_m\fb)\X^{-1}(\nab^n\f)\X^{-1}(\nab_n\fb)\Big]\non\\
	&+\frac{1}{48}\tr\Big[\X^{-1}(\nab^m\f)\X^{-1}(\nab^n\fb)\X^{-1}(\nab_m\f)\X^{-1}(\nab_n\fb)\Big]\non\\
	&+\tr\big[\nab_m \cY^m\big]+\tr\big[\Box \~{\cZ}\big]~.\label{aln:Alttra2(x,x)FinalResult}
\end{align}
Compared to the original field redefinition, the total derivative contributions are the same for $\cY^m$ (\ref{aln:NabTotalDerivativeContribution}) and differ from $\cZ$ (\ref{aln:BoxTotalDerivativeContribution})
\begin{subequations}
	\begin{align}
		\~{\cZ}:=&-\frac{1}{6}\X^{-1}(\nab^m\U)\X^{-1}(\nab_m\U)-\frac{1}{3}\X^{-1}(\Box \X)+\frac{1}{3}\X^{-1}(\nab^m\X)\X^{-1}(\nab_m\X)~.
	\end{align}
\end{subequations}
Indeed the two results for $(a_2)_{\rm total}$ (\ref{aln:tra2(x,x)FinalResult}) and (\ref{aln:Alttra2(x,x)FinalResult}) differ only by a total derivative, which does not contribute to the induced action $\inte^4x\,\sqrt{-g}\, (a_2)_{\rm total}$.

%%%%%%%%%%%%%%%%%%%%%%%%%%%%%
%%%%%%%%%%%%%%%%%%%%%%%%%%%%%

\section{Curved space basis structures}\label{appsect:CurvedSpaceBasisStructures}

Included below is a complete list of the basis structures introduced in (\ref{aln:tra2(x,x)M,NResult}).  Note that under the trace over matrix indices some of these structures are equivalent to one another (via their transpose), however, since these structures are generated directly during the computation we have left them distinct for ease of computational reproducibility.
\begin{align*}
	T_1&:=\X^{-1}(\nab^m\X)\X^{-1}(\nab_m\X)\X^{-1}(\nab^n\X)\X^{-1}(\nab_n\X) ~,\\
	T_2&:=\X^{-1}(\nab^m\X)\X^{-1}(\nab^n\X)\X^{-1}(\nab_m\X)\X^{-1}(\nab_n\X)~,\\
	T_3&:=\X^{-1}(\nab^m\X)\X^{-1}(\nab_m\X)\X^{-1}(\nab^n\U)\X^{-1}(\nab_n\U) ~,\\
	T_4&:=\X^{-1}(\nab^m\X)\X^{-1}(\nab^n\X)\X^{-1}(\nab_m\U)\X^{-1}(\nab_n\U)~,\\
	T_5&:=\X^{-1}(\nab^m\X)\X^{-1}(\nab^n\X)\X^{-1}(\nab_n\U)\X^{-1}(\nab_m\U)~,\\
	T_6&:=\X^{-1}(\nab^m\X)\X^{-1}(\nab_m\U)\X^{-1}(\nab^n\X)\X^{-1}(\nab_n\U)~,\\
	T_7&:=\X^{-1}(\nab^m\X)\X^{-1}(\nab^n\U)\X^{-1}(\nab_m\X)\X^{-1}(\nab_n\U)~,\\
	T_8&:=\X^{-1}(\nab^m\X)\X^{-1}(\nab^n\U)\X^{-1}(\nab_n\X)\X^{-1}(\nab_m\U)~,\\
	T_9&:=\X^{-1}(\nab^m\U)\X^{-1}(\nab_m\U)\X^{-1}(\nab^n\U)\X^{-1}(\nab_n\U)~,\\
	T_{10}&:=\X^{-1}(\nab^m\U)\X^{-1}(\nab^n\U)\X^{-1}(\nab_m\U)\X^{-1}(\nab_n\U)~,\\
	T_{11}&:=\X^{-1}(\Box \X)\X^{-1}(\Box \X)~,\\
	T_{12}&:=\X^{-1}(\Box \U)\X^{-1}(\Box \U)~,\\
	T_{13}&:=\X^{-1}(\Box \X)\X^{-1}(\nab^m\X)\X^{-1}(\nab_m\X)~,\\
	T_{14}&:=\X^{-1}(\Box \X)\X^{-1}(\nab^m\U)\X^{-1}(\nab_m\U)~,\\
	T_{15}&:=\X^{-1}(\Box \U)\X^{-1}(\nab^m\X)\X^{-1}(\nab_m\U)~,\\
	T_{16}&:=\X^{-1}(\Box \U)\X^{-1}(\nab^m\U)\X^{-1}(\nab_m\X)~,\\
	T_{17}&:=R_{mn}\,\X^{-1}(\nab^m\X)\X^{-1}(\nab^n\X)~,\\
	T_{18}&:=R_{mn}\,\X^{-1}(\nab^m\U)\X^{-1}(\nab^n\U)~,\\
	T_{19}&:=R\,\X^{-1}(\nab^m\X)\X^{-1}(\nab_m\X)~,\\
	T_{20}&:=R\,\X^{-1}(\nab^m\U)\X^{-1}(\nab_m\U)~.\\
\end{align*}

%%%%%%%%%%%%%%%%%%%%%%%%%%%%%%%%%%%%%
%%%%%%%%%%%%%%%%%%%%%%%%%%%%%%%%%%%%%%

\begin{footnotesize}

\end{footnotesize}

\end{document}